\documentclass[12pt,english,dvips]{scrartcl}
\usepackage[T1]{fontenc}
\usepackage[ansinew]{inputenc}
\usepackage{amsmath,latexsym,amsfonts}
\usepackage[dvips]{graphicx}
\begin{document}
%\pagecolor{black}
%\color{white}

\begin{titlepage} \vspace{0.2in} 

\begin{center} {\LARGE \bf 
Generating functional for the gravitational field: implementation of an evolutionary quantum dynamics.\\}
\vspace*{0.8cm}
{\bf Erika Cerasti}\\
{\bf Giovanni Montani}\\ \vspace*{1cm}
ICRA---International Center for Relativistic Astrophysics\\ 
Dipartimento di Fisica (G9),\\ 
Universit\`a  di Roma, ``La Sapienza",\\ 
Piazzale Aldo Moro 5, 00185 Rome, Italy.\\ 
e-mail: erika.cerasti@icra.it, montani@icra.it\\ 
\vspace*{1.8cm}

  \vspace*{1cm} \\ 

{\bf Abstract  \\ } \end{center} \indent
We provide a generating functional for the gravitational field, associated to the relaxation of the primary 
constraints as extended to the quantum sector. This requirement of the theory, relies on the assumption that a suitable time variable exist, when taking the T-products of the dynamical variables.\\
More precisely, we start from the gravitational field equations written in the Hamiltonian formalism and expressed via Misner-like variables; hence we construct the equation to which the T-products of the dynamical variables obey and transform this paradigm in terms of the generating functional, as taken on the theory phase-space.\\
We show how the relaxation of the primary constraints (which correspond to break down the invariance of the quantum theory under the 4-diffeomorphisms) is summarized by a free functional taken on the Lagrangian multipliers, accounting  for such constraints in the classical theory.\\
The issue of our analysis is equivalent to a Gupta-Bleuler approach on the quantum implementation of all the gravitational constraints; in fact, in the limit of small $\hbar$, the quantum dynamics is described by a Schr\"odinger equation, as soon as the mean values of the momenta, associated to the lapse function and the shift vector, are not vanishing.\\
Finally we show how, in the classical limit, the evolutionary quantum gravity reduces to General Relativity in the presence of an Eckart fluid, which corresponds to the classical counterpart of the physical clock, introduced in the quantum theory.

\end{titlepage}

\today

\section{Introduction.}
The mean problem we have to face in performing the quantization of the gravitational field is the resulting non evolutionary dynamics, following by the operator translation of the theory constraints. In our points of view, the problem is relied to an ambiguous definition of the label time, in the splitting procedure extended to the quantum regime \cite{Ish}, \cite{tako}, \cite{Kuc3}.\\
Indeed, as shown in \cite{1a}, \cite{1b},  \cite{1c}, \cite{1d}, \cite{1e}, \cite{1f}, \cite{Mon}, a dualism exists in quantum gravity, between including a reference fluid into the dynamics and explicitly breaking down time displacement invariance of the theory. According to this paradigm, we get a self consistent canonical quantization of gravity with the properly evolutionary features, only when a physical reference frame is included into to the system evolution and make possible a physical (3+1)-slicing of the spacetime. In this sense the analysis here presented completes the proof of the dualism inferred in \cite{1d};\\
In fact, we start by assigning a time variable, suitable to construct the T-product functions and show how, in the classical limit of the resulting evolutionary quantum gravity, an Eckart fluid appears.\\
This issue provides a physical meaning to the slicing procedure, in view of the time-like character of the fluid 4-velocity field. Thus, we see how a correct implementation of the classical constraints toward the quantum dynamics, provides a clock for theory and restore the time evolution of the state functional.\\

In section 2 we present the procedure by which a generating functional can be built, in the case of the scalar field,
starting from a field theory approach. The result is achieved using both the Lagrangian and the 
Hamiltonian formalism, after extending the generating functional definition to the phase-space.\\
In section 3, we resume the main steps by which the canonical quantization of gravity is performed, 
i.e. the (3+1)-slicing procedure and its consequences on the dynamics, like the frozen formalism, are discussed.\\
Section 4 deals with the gravitational Hamilton equations, as rewritten in a new set of variables, 
the Misner-like ones; they are introduced in order to make possible the evaluation of a gravitational generating functional $Z$ referred to the phase-space.\\
In the section 5, the procedure illustrated for the scalar field is implemented to the gravitational Hamilton equations and the generating functional is explicitly built up.\\
Through the generating functional formalism, we relax, in the quantum regime, the primary constraints of the gravitational theory, handling on the functional $\mathcal{C}$, taken on the Lagrangian multipliers $\xi$ and $\xi^{i}$. 
In fact, the assumption of a physical time at the ground of the generating functional formalism implies that the corresponding diffeomorphisms invariance is broken, in the quantum gravity approach.
Our point of view is that the problem of the stationary evolution for the gravitational quantum field, stands in a wrong implementation, to the quantum sector, of the classical primary constraints (as well as the implementation of the secondary ones). 
By the constraints relaxation performed in section 6, we obtain, in the limit of small $\hbar$, a Schr\"odinger equation describing the evolution of the quantum gravitational field. We provide the analysis resulting in this equation in section 7. \\
At the end, in section 8, we talk about the classical limit of the founded dynamical equation. This limit accounts for the presence, in the system, of a fluid, having the Eckart form. The presence of this fluid provides the label time $t$ with a physical meaning, that is it assures us about the existence of a non arbitrary time for the considered gravitational system.
In section 9 brief concluding remarks follow.

\section{Generating functional in the scalar field case.} 
As first step, we present the method of searching for the generating functional, used in the scalar field case \cite{Tes};
we show it in both the Lagrangian and Hamiltonian formalism.
Then, in section 5, we will follow this method in order to find the gravitational generating functional expression.
We start to consider a self-interacting scalar field $\varphi(x)$, described by the following action
\[S=\int\left[\,\frac{1}{2}\,\partial_{\nu}\varphi(y)\partial^{\nu}\varphi(y)-
\frac{1}{2}\ \mu^{2}\varphi^{2}(y)-\frac{1}{4}g_{0}\varphi^{4}(y)\right]dy^{4},
\] 
where $\mu$ and $g_{o}$ denote assigned constants;
the associated classical equation of motion read as
\begin{equation}
-\frac{\delta S}{\delta\varphi\left(x\right)}=
\left(\Box_{x}+\mu^{2}\right)\varphi\left(x\right)
+g_{0}\varphi^{3}\left(x\right)=0
\label{eq1moto}
\end{equation}
where $x=(\vec{x},t)$ denotes event coordinates.\\
Now we consider $\varphi$ as a quantum scalar field, satisfying the canonical (equal times) commutation relations,
\begin{eqnarray}
&[\widehat{\varphi}(x),\widehat{\varphi}\left(y\right)]_{x^{0}=y^{0}}&=\;0\nonumber\\
&[\widehat{\pi}(x),\widehat{\pi}\left(y\right)]_{x^{0}=y^{0}}&=\;0\nonumber\\
&[\widehat{\pi}\left(x\right),\widehat{\varphi}\left(y\right)]_{x^{0}=y^{0}}&=\;-i\hbar\,\,\delta^{3}\left(\vec{x}
-\vec{y}\right),
\label{eq1comm}	
\end{eqnarray}
being $\pi(x)\equiv\frac{\partial{\varphi}}{\partial{t}}$ the conjugate momentum to $\varphi(x)$.\\
We define the \emph{two-points T-product} as
\begin{eqnarray}
\left\langle 0\left|T\left(\varphi\left(x\right)\varphi\left(y\right)\right)\right|0\right\rangle &=&
+\,\,\theta\left(x^{0}-y^{0}\right)
\left\langle 0\left|\left(\varphi\left(x\right)\varphi\left(y\right)\right)\right|0\right\rangle\nonumber\\
&& +\,\,\theta\left(y^{0}-x^{0}\right)
\left\langle 0\left|\left(\varphi\left(y\right)\varphi\left(x\right)\right)\right|0\right\rangle\nonumber;
\end{eqnarray}
above, we assumed the existence of a vacuum state for the theory, denoted by $\left|0\right\rangle$. (For a field theory, based on the ground state expectation values, see \cite{Bjd})\\
Let us search for restrictions on the T-products general form, implied by (\ref{eq1moto}), via the commutation relations (\ref{eq1comm}). 
Starting with the expectation value of the classical equation, taken on the ground state,
\begin{eqnarray*}
\left\langle 0\left|T\left(\left[\Box_{x}\varphi\left(x\right)+\mu^{2}\varphi\left(x\right)
+g_{0}\varphi^{3}\left(x\right)\right]\varphi\left(y\right)\right)	\right|0\right\rangle=0,
\end{eqnarray*}
we arrive to an equation for the two-points T-products
\[\left[\Box_{x}+\mu^{2}\right]\langle0\left|T\left(\varphi\left(x\right)\varphi\left(y\right)\right)\right|0\rangle
+g_{0}\left\langle 0\left|T\left(\varphi^{3}\left(x\right)\varphi\left(y\right)\right)\right|0\right\rangle=-
i\hbar\delta^{4}\left(x-y\right)
\]
(For a T-products theory see \cite{Bjd}).\\
This equation can be generalized in view of n-points T-products as follows:
\begin{eqnarray}
&&\left[\Box_{x} + \mu^{2}\right]\langle0\left|T\left(\varphi\left(x\right)\varphi\left(x_{1}\right)...
\phi\left(x_{n}\right)\right)\right|0\rangle+
g_{0}\left\langle 0\left|T\left(\varphi^{3}\left(x\right)\varphi\left(x_{1}\right)...
\varphi\left(x_{n}\right)\right)\right|0\right\rangle\nonumber\\ &&=-i\hbar\sum^{n}_{i=1}\delta^{4}\left(x-x_{i}\right)\left\langle 0\left|T\left(\varphi\left(x_{1}\right)...\varphi\left(x_{i-1}\right)\varphi\left(x_{i+1}\right)...
\varphi\left(x_{n}\right)\right)\right|0\right\rangle \label{eqi1nmoto}
\end{eqnarray}
Introducing the generating functional $Z(J)$ (\cite{Ram}, \cite{ZJ}, \cite{IZ}) is useful to rewrite 
the whole set of the equation (\ref{eqi1nmoto}), namely for any value of n, in a compact form.\\
We assign the source current $J(x)$ and let us construct the quantity
\[Z(J)=\sum^{\infty}_{n=0}\frac{i^{n}}{n!}\int dx^{4}_{1}...dx^{4}_{n}
\left\langle 0\left|T\left(\varphi\left(x_{1}\right)...\varphi\left(x_{n}\right)\right)\right|0\right\rangle
J\left(x_{1}\right)...J\left(x_{n}\right),
\]
which satisfies the normalization condition $Z(0)=1$; \\
in a shorter form, $Z$ can be restated in terms of the following vacuum expectation value
\[Z\left(J\right)=\left\langle 0\left|T\left(e^{i\int J\varphi}\right)\right|0\right\rangle,
\]
where we have:
$$
i\int J\varphi=i\int J\left(x\right)\varphi\left(x\right)dx^{4}.
$$
We show explicitly the link between the T-product functions and the functional derivatives of $Z$ as
\[\left\langle 0\left|T\left(\varphi\left(x_{1}\right)...\varphi\left(x_{n}\right)\right)\right|0\right\rangle=
\frac{1}{i^{n}}\frac{\delta^{n}Z\left(J\right)}{\delta J\left(x_{n}\right)...\delta J\left(x_{1}\right)}\Bigg{|}_{J=0};
\]
so we can get the equation for any T-product by performing the derivatives of the following fundamental expression
\begin{equation}
\left[\Box_{x}+\mu^{2}\right]\frac{1}{i}\frac{\delta Z\left(J\right)}{\delta J\left(x\right)}+
g_{0}\frac{1}{i^{3}}\frac{\delta^{3} Z\left(J\right)}{\delta^{3} J\left(x\right)}=
\hbar J\left(x\right)Z\left(J\right)\label{basic}.
\end{equation}
To solve (\ref{basic}), we introduce $\widetilde{Z}(\varphi)$, the functional Fourier transformation of $Z(J)$, 
\[Z\left(J\right)=\int\prod_{x}d\varphi(x)\widetilde{Z}(\varphi)
e^{i\int\varphi(x)J(x)dx^{4}}\]
In the next, we set the Lebesgue measure as $\delta\varphi=\prod_{x}d\varphi\left(x\right)$.\\
If we substitute, into (\ref{basic}), the Fourier expansion of $Z$, we get the equation
\[\int\delta\varphi e^{i\int\varphi J}\left\{\widetilde{Z}\left(\varphi\right)
\left[\left(\Box_{x}+\mu^{2}\right)\varphi\left(x\right)+g_{0}\varphi^{3}\left(x\right)\right]+
\frac{\hbar}{i}\frac{\delta\widetilde{Z}\left(\varphi\right)}{\delta\varphi\left(x\right)}\right\}=0.
\]
With some algebra and using (\ref{eq1moto}), we arrive to the expression
\begin{eqnarray*}
\frac{\delta}{\delta\varphi\left(x\right)}\left[ln\widetilde{Z}(\varphi)-\frac{i}{\hbar}S(\varphi)\right]=0,
\end{eqnarray*}
by which we finally recognize the functional $\widetilde{Z}$ in the form
\[\widetilde{Z}(J)=C e^{\,\frac{i}{\hbar}S\left(\varphi\right)}, 
\]
so that the generating functional follows:
\[Z\left(J\right)=C\int\delta\varphi(x)\;e^{\,\frac{i}{\hbar}S\left(\varphi\right)+ i\int J(x)\varphi(x)d^{4}x}, 
\]
being $C$ the normalization constant.\\
In view of applying this procedure to the gravitational case, we have to translate it, into the Hamiltonian formalism, because of some difficulties that Lagrangian formalism can not overcome.
So, we get the Hamiltonian of the system out of the scalar field action,
$$
H=\int\left[\frac{1}{2}\,\pi^{2}-\frac{1}{2}\,\partial^{j}\varphi\,\partial_{j}\varphi+\frac{1}{2}\ \mu^{2}\varphi^{2}+\frac{1}{4}\,g_{0}\varphi^{4}\right]dy^{3},
$$ 
via the above conjugate momentum of $\varphi(x)$.\\
The Hamiltonian equations we find, look as follows:
\begin{eqnarray}
\dot{\pi}&=&-\frac{\delta H}{\delta\varphi}=\left(\partial^{2}_{j}\varphi-\mu^{2}\varphi\right)
-g_{0}\varphi^{3}\label{ham1}\\[0.2cm]
\dot{\varphi}&=&+\frac{\delta H}{\delta\pi}=\pi.\label{ham2}
\end{eqnarray}
Like before, we transform each equation of motion in a T-products one, using again the commutation relations (\ref{eq1comm})
\[\left\{\begin{array}{lll}
\partial_{t}\left\langle 0\left|T\left[\pi(x)\varphi(y)\right]\right|0\right\rangle
&+&\partial^{2}_{j}\left\langle 0\left|T\left[\varphi(x)\varphi(y)\right]\right|0\right\rangle
+\mu^{2}\left\langle 0\left|T\left[\varphi(x)\varphi(y)\right]\right|0\right\rangle\\[0.4cm]
&+&g_{0}\left\langle 0\left|T\left[\varphi^{3}(x)\varphi(y)\right]\right|0\right\rangle
=-i\hbar\delta^{4}(x-y)\\[0.4cm]
\partial_{t}\left\langle 0\left|T\left[\varphi(x)\pi(y)\right]\right|0\right\rangle
&-&\,\,\,\,\,\left\langle 0\left|\,T\left[\,\pi(x)\pi(y)\,\right]\right|0\right\rangle\,\,
=\,\,+i\hbar\,\,\delta^{4}(x-y).
\end{array}\right.
\]
Let us provide the definition of the generating functional, extended to the phase-space
\[Z\left(J,W\right)=\left\langle 0\left|T\left(e^{i\int(J\varphi+W\pi)}\right)\right|0\right\rangle
\]
and its Fourier Transform
\begin{equation}
Z\left(J,W\right)=\int\delta\varphi\,\delta\pi\,\widetilde{Z}\left(\varphi,\pi\right)\;
e^{i\int \left[J(x)\varphi(x)+W(x)\pi(x)\right]dx^{4}}
\label{fourier}
\end{equation}
where $W$ is the current source associated to the conjugate momentum.\\
Thus, the equations of motion become functional ones in $Z(J,W)$
\begin{equation}
\partial_{t}\,\frac{1}{i}\frac{\delta Z}{\delta W(x)}+\left[\partial^{2}_{j}+\mu^{2}\right]\,\frac{1}{i}
\frac{\delta Z}{\delta J(x)}+g_{0}\,\frac{1}{i^{3}}\frac{\delta^{3}Z}{\delta J(x)^{3}}
=\hbar\,Z(\varphi)J(x)\label{fond1}
\end{equation}
\begin{equation}
\partial_{t}\,\frac{1}{i}\frac{\delta Z}{\delta J(x)}+\frac{1}{i}\frac{\delta Z}{\delta W(x)}+
=-\hbar\,Z(\varphi)W(x). \label{fond2}
\end{equation}
Hence, the first of these equations can be rewritten in terms of $\widetilde{Z}(J,W)$ and takes the form
\begin{eqnarray*}
\int\delta\varphi\delta\pi\widetilde{Z}\left(\varphi,\pi\right)e^{i\int(J\varphi+W\pi)}&
\left[\partial_{t}\pi(x)-\left(\partial_{j}+\mu^{2}\right)\varphi(x)+g_{0}\varphi^{3}(x)\right]\nonumber\\[0.2cm] 
&=-\frac{\hbar}{i}\int\delta\varphi\delta\pi\;\left(\frac{\delta\widetilde{Z}\left(\varphi,\pi\right)}
{\delta\varphi\left(x\right)}\right)\;e^{i\int(J\varphi+W\pi)}.
\end{eqnarray*}
At the end, with simple algebra, we get
\[\frac{\delta}{\delta\varphi\left(x\right)}
\left[ln\widetilde{Z}(\varphi,\pi)-\frac{i}{\hbar}S(\varphi,\pi)\right]=0,
\]
which admits the solution
\begin{equation}
\widetilde{Z}\left(\varphi,\pi\right)=C(\pi)\,e^{\,\frac{i}{\hbar}S\left(\varphi,\,\pi\right)}\label{til1};
\end{equation}
here, $C(\pi)$ denotes an integration $\varphi$-constant functional.
In analogy from (\ref{fond2}), we get also
\begin{equation}
\widetilde{Z}(\varphi,\pi)=D(\varphi)\,e^{\,\frac{i}{\hbar}\,S(\varphi,\pi)}\label{til2}
\end{equation}
being $D(\varphi)$ an integration $\pi$-constant functional.
Compatibility of (\ref{til1}), (\ref{til2}) requires 
that the following expression for $Z(J,W)$ holds
\[Z(J,W)=N\int\delta\pi\delta\varphi\;e^{\,\frac{i}{\hbar}S\left(\varphi,\pi\right)+ i\int (J\varphi+W\pi)} 
\]
where $N$ is pure normalization constant. 
The above is just the procedure which will allow us to get.\\
We conclude this section by remarking that, in the scalar field case, the Hamiltonian formulation of the 
generating functional is equivalent to the Lagrangian one, as soon as we address the identifications $\pi=\partial_{t}\varphi$ and $W\equiv0$.

\section{Canonical quantization of gravity in the frozen formalism.}
Before we begin to search for the gravitational generating functional, we resume the necessary steps allowing us to arrive to the canonical quantum equation for the gravitational field, that is the Wheeler-DeWitt equation \cite{Kuc2}, \cite{Dew}, \cite{DewG};\\
We start to consider a pseudo-Riemannian 4-dimensional manifold $M^{4}$, on which a metric tensor $h_{\mu\nu}(y^{\rho})$ is defined. 
The signature we choose is $(- + + +)$. (The Greek indexes stand for 0,1,2,3, as the Latin ones stand for 1,2,3).\\ 
We apply a (3+1)-slicing assigning the parametric equation of a space-like hypersurfaces family $\sum^{3}_{t}:y^{\rho}=y^{\rho}(t,x^{k})$ on the considered manifold.\\ 
This definition provides a local reference basis $\{e^{\mu}_{i},n^{\mu}\}$ in $M^{4}$, made up by the tangent vectors $e^{\mu}_{i}\equiv\partial_{i}y^{\mu}$ and the normal vector $n^{\mu}(y^{\rho})$ to $\sum^{3}_{t}$. 
Projecting the \emph{deformation vector} $N^{\mu}$ on this basis we get the \emph{lapse function} $N(t,x^{k})$ and the \emph{shift vector} $N^{i}(t,x^{k})$ \cite{ADM4}:
\begin{equation} 
N^{\mu}\equiv \partial_ty^{\mu}=Nn^{\mu}+N^i
e^{\mu}_i.
\label{defo} 
\end{equation} 
In terms of this quantity we find the metric tensor expression in the $(t,x^{k})$-reference system
\begin{equation}
h_{\mu\nu}=\left(
\begin{array}
[c]{cc}%
N_{i}N^{i}-N^{2} & N_{i}\\
N_{i} & g_{ij}%
\end{array}
\right). 
\end{equation}
The term $g_{ij}\equiv h_{\mu\nu}\,e^{\mu}_{i}\,e^{\nu}_{j}$ is the so-called induced metric tensor, i.e. the projection of the metric tensor on the space-like hypersurface, which becomes our dynamical variable.\\
Now the splitted Hilbert-Einstein action (the A.D.M. action) looks like as the following one \cite{ADM4}, \cite{ADM1}, \cite{ADM2}, \cite{ADM3}, \cite{Thi}, \cite{MTW}, \cite{Wal}:
\begin{equation}
S_{HE}=\underset{\Sigma^{3}\times\Re}
{\displaystyle\int}
dtd^{3}xN\sqrt{g}\left(R+k_{ij}k^{ij}-k^{2}\right)  ,
\label{a.gravitazionale}
\end{equation}
where $k_{ij}$ denotes the extrinsic curvature of the tre-hypersurfaces $\Sigma^{3}$, while $R$ is the curvature scalar associated to the tre-metric.\\
From the Lagrangian contained in the previous expression, we find the classical momenta associated to $g_{ij}$ and to $N$ and $N^{i}$, in view of rewriting the action in the Hamiltonian formalism. After the Legendre transformation, the action rewrite as
\begin{equation}   
S_{HE} = \int _{\Sigma^{3}\times\Re} \left\{\pi ^{ij}\partial _tg_{ij}
- N\mathcal{H} - N^i\mathcal{H}_{i}\right\} d^3x\,dt  
\label{s} 
\end{equation}  
where $\mathcal{H}$ and $\mathcal{H}_{i}$ are called super-Hamiltonian and super-momentum respectively.
The momenta $P_{N}$ and $P_{N^{i}}$, associated to the lapse function and to the shift vector, vanish identically, as
\begin{equation}
P_{N}=\frac{\delta L}{\delta\left(
\partial_{t}N\right)}=0,\qquad P_{N^{i}}=\dfrac
{\delta L}{\delta(\partial_{t}N^{i})}=0
\end{equation} 
and these equations constitute the primary constraints for the gravity dynamics.
The secondary constraints follow directly from the first ones, via the Hamilton equations (see also \cite{Kuc2}, \cite{ADM4}, \cite{Thi}):
\begin{eqnarray}
\mathcal{H}(g_{ij},\pi^{ij})&=&G_{ijkl}\pi^{ij}\pi^{kl} -
\sqrt{h}{}^3R=0\,, \\[0.2cm]
\mathcal{H}_{i}(g_{ij},\pi^{ij})&=& -2{}^3\nabla _j\pi ^j_i=0\,,\label{sec}
\end{eqnarray}
where $G_{ijkl}$ is the super-metric and stands as
\begin{equation*}
G_{ijkl} \equiv \frac{1}{2\sqrt{h}}
(h_{ik}h_{jl} + h_{il}h_{jk} - h_{ij}h_{k})\,.  
\end{equation*}
The canonical quantization procedure is obtained by translating secondary constraints
into operator ones, on the wave functional $\psi$, which describes the physical states, i.e.
\begin{equation*}
\widehat{\mathcal{H}}(\widehat{g}_{ij},\widehat{\pi}^{ij})\,\psi=0\,,
\label{vincolo op1}
\end{equation*}
\begin{equation*}
\widehat{\mathcal{H}}_{i}(\widehat{g}_{ij},\widehat{\pi}^{ij})\,\psi=0\,,
\label{vincolo op2}
\end{equation*}
where we introduced the canonical quantum operators $\widehat{g}_{ij},\,\widehat{\pi}^{ij}$, for which the commutation relations
\begin{eqnarray}
\left[\,\widehat{g}_{ab}(\vec{x},t)  ,\widehat{g}_{cd}
(\vec{x}\,^{\prime},t)  \right]  &=&0,\nonumber\\[0.2cm]
\left[  \widehat{\pi}\,^{ab}(\vec{x},t)  ,\widehat{\pi}\,^{cd}
(\vec{x}\,^{\prime},t)  \right]  &=&0,\nonumber\\[0.2cm] 
\left[  \widehat{g}_{ab}\left(  \vec{x},t\right)  ,\widehat{\pi}\,^{cd}
(\vec{x}\,^{\prime},t)  \right]  &=&i\hbar\,\delta_{ab}^{cd}\,\delta
(\vec{x}-\vec{x}\,^{\prime})   \label{comm3}
\end{eqnarray}
have to be valid; this last request implies the following representation for the tre-metric and momenta:
\begin{equation*}
\widehat{g}_{ij}(x)\,\psi  
=g_{ij}(x)\psi\,,
\end{equation*}
$\qquad$
\begin{equation*}
\widehat{\pi}^{kl}(x)\,\psi
=-i\hbar\dfrac{\delta\psi}{\delta g_{kl}(x)}\,.
\end{equation*}
At the end, choosing the appropriate normal ordering \cite{Kuc1}, we can rewrite the operator constraints as
\begin{equation*}
\widehat{\mathcal{H}}(x)\,\psi=\left[-\hbar^{2}:G_{ijkl}(x)
\dfrac{\delta^{2}\psi}{\delta g_{ij}(x)\delta g_{kl}(x)}
:-\sqrt{g}R\right]\,\psi=0\,,
\end{equation*}

\begin{equation*}
\widehat{\mathcal{H}}_{i}(x)\,\psi=:2i\,\hbar\,g_{ik}\nabla_{j}\dfrac
{\delta\psi}{\delta g_{jk}(x)}:=0\,.
\end{equation*}
The first of the above functional equations is known as the Wheeler-DeWitt one \cite{Dew} and provides the quantum dynamics of the system, while the tree super-momentum equations state the theory invariance under tre-diffeomorphisms. This last invariance provides the wave functional depending only on tre-geometries $\{g_{ij}\}$, i.e. $\psi=\psi(\{g_{ab}\})$ \cite{Wal}, \cite{Kuc1}, \cite{KT}.\\
As we can see, the Wheeler-DeWitt equation is a Sch\"orodinger equation for a non-dynamical system, characterized by a wave functional satisfying $\partial_{t}\psi=0$;
In detail, the $N$-independence of $\psi$ is responsible for the time non-evolution of the wave functional.\\
The point of view addressed in this paper is that such stationary evolution for the gravitational field is caused by a too straightforward quantum implementation of the theory constraints, in the canonical form. In particular, we concentrate our attention on the quantum formulation of the primary constraints; in the canonical Wheeler-DeWitt approach, the translation of these constraints, $P_{N}=0$ and $P_{N^{i}}=0$, leads to the independence of the wave functional on the lapse function and on the shift vector, i.e.
\begin{eqnarray*}
\widehat{P}_{N}\,\psi=-i\hbar\frac{\delta}{\delta N}\,\psi=0\,,\\[0.2cm]
\widehat{P}_{N^{i}}\,\psi=-i\hbar\frac{\delta}{\delta N^{i}}\,\psi=0\,,
\end{eqnarray*}
an issue consistent with the non-evolutionary character of the WDE along the slicing and of its invariance under the tre-diffeomorphisms.\\
In the next section, in the spirit of the Gupta-Bleuler approach \cite{Mdls}, we will propose a relaxation of these constraints, allowing an evolution for the wave functional $\psi(g_{ij},t)$, which depends so on time and on the tre-metric field.\\

\section{The gravitational theory in the Misner-like variables.}
We start by introducing the Misner-like variables $(\alpha(x),u(x))$ \cite{Mis}, via the relations
\begin{eqnarray}
g_{ij}(x)\equiv e^{\;\;\alpha(x)}\left(e\;\;^{u(x)}\right)_{ij}\\[0.2cm]
g^{ij}(x)\equiv e^{-\alpha(x)}\left(e^{-u(x)}\right)^{ij}\,,\label{met}
\end{eqnarray}
where $u(x)$ is a traceless matrix and, in order to safe the tensorial language, we assign a different character to the quantity $e^{u(x)}$ and $u(x)$ itself, i.e. covariant and mist one respectively.\\
The aim of the below analysis is to rewrite the gravitational Hamiltonian $H(g_{ij},\pi^{ij})$ in the Misner-like variables, so then we can get the equations of motion in $(\alpha(x),u(x))$ and use them to find the generating functional for gravity, like shown in the previous section.\\  
The relation between the old momenta $\pi^{ij}$, and the new ones is provided by requiring the canonical nature of the transformation, i.e.
\begin{equation}
\pi^{ij}\,\partial_{t}g_{ij}=p_{\alpha}\,\partial_{t}\alpha+ p^{k}_{i}\,\partial_{t}u^{i}_{k}\,.
\label{tracano}\end{equation}
Substituting the relation (\ref{met}) into (\ref{tracano}) we can identify the new momenta as
\[\left\{\begin{array}{ll}
p_{\alpha}=\mathbf{\Pi}\equiv\pi^{ij}g_{ij}\\[0.2cm]
p^{k}_{i}=\pi\,^{k}_{i}-\frac{1}{3}\,\mathbf{\Pi}\,\delta^{k}_{i}\,,
\end{array}\right.	
\]
where $\mathbf{\Pi}$ denotes the quantity $\pi^{ij}g_{ij}$.
Starting from the expression of the super-Hamiltonian $\mathcal{H}$ and super-momentum $\mathcal{H}_{i}$, 
we rewrite them in terms of the new variables as
\[\left\{\begin{array}{ll}
\mathcal{H}=&\frac{1}{2}\,e^{-(3/2)\alpha}\left[\,2\,p^{j}_{i}\,p^{i}_{j}-\frac{1}{3}\,p_{\alpha}\,^{2}\right]
-e^{(3/2)\alpha} R(u,\alpha)\nonumber\\[0.3cm]
\mathcal{H}_{i}=&-2\,\partial_{j}\,p\,^{j}_{i}-\frac{2}{3}\,\partial_{i}\,p_{\alpha}
+p_{\alpha}\,\partial_{i}\alpha+2\,U^{j}_{ni}p^{n}_{j}\,,\nonumber
\end{array}\right.
\]
where we denote with $U^{n}_{dq}(u,\partial u)$ the Christoffel terms constructed only by derivatives of $u$-variable,
\[\left\{\begin{array}{lll}
U^{n}_{dq}(u,\partial u) &\equiv& \frac{1}{4}\Big[+\partial_{d}u^{n}_{q}+\partial_{q}u^{n}_{d}+\partial_{q}u^{m}_{l}(e^{-u})^{nl}
(e^{u})_{md}+(\partial_{d}u^{m}_{l})(e^{-u})^{nl}(e^{u})_{mq}\\[0.2cm]
&&-(\partial_{l}u^{m}_{d})(e^{-u})^{nl}(e^{u})_{mq}-(\partial_{l}u^{m}_{q})(e^{-u})^{nl}(e^{u})_{md}\Big]\,,
\end{array}\right.
\]
and the full Christoffel symbol rewrites as
\[\Gamma^{n}_{dq}\equiv U^{n}_{dq}+\frac{1}{2}\Big[(\partial_{d}\alpha)\delta^{n}_{q}+(\partial_{q}\alpha)\delta^{n}_{d}-
(\partial_{m}\alpha)(e^{-u})^{nm}(e^{u})_{dq}\Big]\,.
\]
Starting from the expression of $\mathcal{H}$ and $\mathcal{H}_{i}$, as written in the new variables, we can get the Hamilton equations for the gravitational system, in the Misner-like representation, which stand as

\begin{eqnarray*}
\dot{u}^{b}_{a}&=&- 2Ne^{-(3/2)\alpha}p\,^{b}_{a}
-2 \left[U^{a}_{bn}N^{n}+\partial_{a}N^{b}\right]
\\[0.4cm]
\dot{p}^{b}_{a}&=&-\Theta^{b}_{a}+2 
\,\Omega\,^{ib}_{jna}N^{n}p^{j}_{i}
\\[0.4cm]
\dot{\alpha}&=&+\frac{1}{3}\,Ne^{-(3/2)\alpha}\,p_{\alpha}
-\left[(\partial_{l}\alpha)N^{l}+\frac{2}{3}\,\partial_{j}N^{j}\right]\\[0.4cm]
\dot{p_{\alpha}}&=&-Ne^{(3/2)\alpha}\left\{\frac{3}{4}\,e^{-3\alpha}
\left[+2\,p\,^{j}_{i}\,p\,^{i}_{j}-\frac{1}{3}\,p_{\alpha}^{2}\right]
+\frac{3}{2}\,R(u,\alpha)\right\}-\Theta\\[0.2cm]
&&-\partial_{l}\left(N^{l}p_{\alpha}\right)
\end{eqnarray*}
being
\begin{equation*}
\int d^{3}x\,Ne^{(3/2)\alpha}\left(\frac{\delta R(u,\alpha)}{\delta\alpha}\right)\delta\alpha
\equiv\int d^{3}x\,\Theta\,\delta\alpha
\end{equation*}
\begin{equation*}
\int d^{3}x\,Ne^{(3/2)\alpha}\left(\frac{\delta R(u,\alpha)}{\delta u^{a}_{b}}\right)\delta u^{a}_{b}
\equiv\int d^{3}x\,\Theta^{b}_{a}\,\delta u^{a}_{b}
\end{equation*}
and 
\begin{eqnarray*}
\int d^{3}x \left(\frac{\delta U\,^{i}_{j\,n}}{\delta u^{a}_{b}}\right)N^{n}\,\,\delta u^{a}_{b}\equiv\int d^{3}x\,\Omega\,^{ib}_{jna}\,\delta u^{a}_{b}\,.
\end{eqnarray*}
The introduction of these new variables, together with the Hamiltonian approach, is necessary to overcome some 
difficulties in applying the method, shown in section 2, to the gravitational case; such difficulties 
can be summarized by the following point:\\
\begin{itemize}
\item the presence of the inverse metric tensor in the equations of motion, which we can not easily 
rewrite as a function of the direct metric $g_{ij}$; now, both the inverse and direct metric are function of the dynamical variables, $(\alpha,u)$, and so no difficulty survives.\\
\item the impossibility to take all the temporal derivatives out of the field T-products, because of their non linearity;
using an Hamiltonian approach, quadratic terms containing time derivatives of the tre-metric tensor are removed, because we rewrite them via the conjugate momenta. In this sense, the phase-space representation overcomes the gravity non-linearity problem in building up the generating functional (the non-linearity in the spatial derivatives plays no role here).
\end{itemize}

\section{Gravitational generating Functional.}
Using the Misner-like variables and the Hamiltonian formulation, the Hilbert-Einstein action
looks like 
\begin{eqnarray*}
S=\int_{M^{4}} d^{3}xdt&\Big[&p^{j}_{n}\;\partial_{t}u^{n}_{j}+p_{\alpha}\partial_{t}\alpha+ P_{N}\partial_{t}N+
P_{N^{i}}\partial_{t}N^{i}\\
&&-\xi P_{N}-\xi^{i} P_{N^{i}}- N\mathcal{H} - N^{i}\mathcal{H}_{i}\,\,\,\,\,\Big]\,,
\end{eqnarray*}
where the Lagrange multipliers $\xi(x),\xi^{i}(x)$ are included in order to obtain the primary 
constraints ($P_{N}\equiv P_{N^i}=0$) as Hamilton equations, and
being the integration domain ${M^{4}}=\Sigma^{3}\times\mathcal{R}$ ($\Sigma^{3}$ is taken compact and boundaryless).\\
The whole set of Hamilton equations is, therefore,
\[\begin{array}{lrcl}
\mbox{I)} & +\dot{u}\,^{n}_{j}-\frac{\delta H}{\delta p\,^{j}_{n}}&=&0\\[0.3cm]
\mbox{II)} & -\dot{p}\,^{n}_{j}-\frac{\delta H}{\delta u\,^{j}_{n}}&=&0\\[0.3cm]
\mbox{III)} & +\dot{\alpha}-\frac{\delta H}{\delta p_{\alpha}}&=&0\\[0.3cm]
\mbox{IV)} & -\dot{p}_{\alpha}-\frac{\delta H}{\delta \alpha}&=&0\\[0.3cm]
\mbox{V)} & -\dot{P}_{N}- \mathcal{H}&=&0\\[0.3cm]
\mbox{VI)} & -\dot{P}_{N^{i}}- \mathcal{H}_{i}&=&0\\[0.3cm]
\mbox{VII)} & +\dot{N}-\xi &=&0\\[0.3cm]
\mbox{VIII)} & +\dot{N^{i}}-\xi^{i}&=&0\\[0.3cm]
\mbox{IX)}& -P_{N}&=&0\\[0.3cm]
\mbox{X)}& -P_{N^{i}}&=&0\,.
\end{array}
\]
In order to extend the analysis of section 2 to the gravitational case, we assume that the following statements are valid for the quantum theory:
\begin{itemize}
\item We can define a ground (vacuum) state $\left|0\right\rangle$, for the gravitational theory, by which we can built the T-product of the operator field and of its momentum. For the same reason, we need the theory is provided with a definite time variables, which has to have the meaning of physical time; we identify it with the label time.
\item By a second quantization procedure $g_{ij}(x)$ and $\pi^{ij}(x)$ become operators which satisfy the canonical commutation relations (\ref{comm3}). This hypothesis has to hold for any representation of the tre-metric and of the conjugate momentum, like the considered Misner-like one.
\end{itemize}
Now, we introduce the generating functional $Z$ \cite{ZJ}, \cite{IZ},
\begin{eqnarray}
Z(J^{a},W_{a})\equiv\left\langle 0\left|T\left(e^{i\int\left(g_{a}J^{a}+P^{a}W_{a}\right)}\right)
\right|0\right\rangle\,,
\label{zeta}
\end{eqnarray}
where we adopt a notation of the form
\[Z\left(J,W\right)= Z\left(J^{j}_{i},J_{\alpha},J_{N},J_{N^{i}},W^{i}_{j},W_{\alpha},W_{N},W_{N^{i}}\right)\,,
\]
\begin{eqnarray}
\int\left(g_{a}J^{a}+P^{a}W_{a}\right)&=& \int \Big[ u^{n}_{j}(x)J^{j}_{n}(x)+p^{j}_{n}(x)W^{n}_{j}(x)+ \alpha(x) J_{\alpha}(x)\nonumber\\[0.2cm]
&&+ p_{\alpha}(x)W_{\alpha}(x)+N(x)J_{N}(x)+  N^{i}(x)J_{N^{i}}(x)\nonumber\\[0.2cm]
&&+ P_{N}(x)W_{N}(x)+P_{N^{i}}(x)W_{N^{i}}(x)\Big]d^{3}x \,dt\,.\nonumber
\end{eqnarray}
As in the scalar case, even here $J^{a}$ and $W_{a}$ have to be interpreted as quantum source currents, because we will see they give a $\hbar$ contribution in the action.\\
Now, we integrate the equations I)-VIII) to determine the generating functional, in the gravitational theory, via an algorithm which we discuss here, in detail, only for equation I).\\

From the first equation, we build the corresponding T-products dynamics
\[\partial_{t}\left\langle 0\right|T\left(u^{n}_{j}\left(x\right)p^{j}_{n}\left(y\right)\right)\left|0\right\rangle-
\left\langle 0\right|T\left(\frac{\delta H}{\delta p^{j}_{n}\left(x\right)}\;\; p^{j}_{n}\left(y\right)\right)\left|0\right\rangle=
i\hbar\;\delta^{4}\left(x-y\right)\,,
\]
which can be rewritten formally in terms of $Z$, as
\begin{equation}
\partial_{t}\;\frac{1}{i}\frac{\delta Z}{\delta J^{j}_{n}(x)}
-H \left(-i\frac{\delta} {\delta J^{a}}\,,-i\frac{\delta }{\delta W_{a}}\right)Z =-\hbar\,W^{n}_{j}(x)\,Z\,.
\label{fondamentale}
\end{equation}
Substituting, in the previous equation, the following Fourier Transform
\begin{equation}
Z\left(J,W\right)=\int D(g_{a},P^{a})\,\widetilde{Z}(g_{a},P^{a})\,e^{i\int\left(g_{a}J^{a}+P^{a}W_{a}\right)}\,,
\label{TF}
\end{equation}
where the Lebesgue measure $D(g_{a},P^{a})$ is
\[D(g_{a},P^{a})=\delta u^{n}_{j}\,\delta\,\alpha\,\delta N\,\delta N^{i}\,\delta p^{j}_{n}\,\delta P_{\alpha}\,\delta P_{N}\,\delta P_{N^{i}}\,,
\]
we get  
\[\int D(g_{a},P^{a})e^{i\int\left(g_{a}J^{a}+P^{a}W_{a}\right)}\,\widetilde{Z}\left(\dot{u}^{n}_{j}
-\frac{\delta H}{\delta p^{j}_{n}}\right)=
\int D(g_{a},P^{a})e^{i\int\left(g_{a}J^{a}+P^{a}W_{a}\right)}\,\frac{\hbar}{i}\frac{\delta\widetilde{Z}}{\delta p^{j}_{n}}\,.
\]
This functional equation admits the solution
\[\widetilde{Z}(g_{a},P^{a})=C\left(-p\,^{j}_{n}-\right)\; e^{\frac{i}{\hbar}S}\,.
\]
$C\left(-p\,^{j}_{n}-\right)$ denotes a functional depending on all the variables (including $\xi$ and $\xi^{i}$) but $p\,^{j}_{n}$.\\
Inferring the above procedure for all the other equations, but the last two, we arrive to a final expression for $Z$, i.e.
\[Z\left(J^{a},W_{a}\right)=\mathcal{I}\int D(g_{a},P^{a})\delta\xi \delta\xi^{i}\,\mathcal{C}\left(\xi,\xi^{i}\right)\; e^{\frac{i}{\hbar}S+i\int\left(g_{a}J^{a}+P^{a}W_{a}\right)}\,,
\]\\
being $\mathcal{I}$ a normalization constant and
$\mathcal{C}\left(\xi,\xi^{i}\right)$ an arbitrary functional on the Lagrangian multipliers, allowed by the theory; we remark that the integration over such multipliers is required by the $Z$ independence on them (the T-products calculated by $Z$ are observable of the theory and, therefore, can not depend on Lagrangian multipliers), and it is possible in view of the linearity, characterizing the whole system of equation I)-VIII).\\
The implementation of the primary constraints IX)-X), via the above algorithm, would imply the $Z$ independence on $W_{N}$ and $W_{N^{i}}$, which would lead again to the Wheeler-DeWitt dynamics; however, our aim is to investigate the quantum dynamics resulting from different implementation of this primary constraints. As we shall see in the next section, the role played here by equations IX)-X), in the quantum sector, is resumed by the form of the functional $\mathcal{C}\left(\xi,\xi^{i}\right)$. (In fact, the existence of this arbitrariness in the theory is a direct consequence of disregarding the primary constraints).

\section{The Gupta-Bleuler approach.}
We have ended the previous section, saying the form we choose for the functional $\mathcal{C}(\xi,\xi^{i})$ reflects the way by which we perform the implementation of the primary constraints in the quantum theory.\\
We start by observing how the frozen formalism requirement, that the operators $\widehat{P}_{N}$ and $\widehat{P}_{N^{i}}$ annihilate the Sch\"orodinger wave functional, is equivalent, in the Heisenberg approach, to have vanishing T-products of these operators; \\
indeed, the vanishing behavior of T-products, of any number of points, implies that the generating functional does not depend on the currents $W_{N}$ and $W_{N^{i}}$, i.e. 
\[-i\frac{\delta Z}{\delta W_{N}}=0\,,  
\]
\[-i\frac{\delta Z}{\delta W_{N^{i}}}=0\,.  
\]
Below, we show that the  quantum implementation of the primary constraints is summarized by the functional form taken by $\mathcal{C}(\xi,\xi^{i})$, and, therefore, this quantity is the fundamental degree of freedom, in our approach.\\
We start by observing that the resulting generating functional  
\begin{eqnarray*}
Z\left(J,W\right)=&\mathcal{I}&\int D(g_{a},P^{a})\,\delta\xi\,\delta\xi^{i}\, 
\mathcal{C}\left(\xi,\xi^{i}\right)\\[0.2cm]
&&\qquad e^{\frac{i}{\hbar}S+\int \Big[ u^{n}_{j}J^{j}_{n}+p^{j}_{n}W^{n}_{j}+ \alpha J_{\alpha}
+ p_{\alpha}W_{\alpha}+NJ_{N}+ N^{i}J_{N^{i}}\Big]d^{3}x \,dt}\,,
\end{eqnarray*}
contains integrals of the form
\[\int \delta\xi(\vec{x},t)\, \delta\xi^{i}(\vec{x},t)\,\mathcal{K}(\xi)\mathcal{K}_{i}(\xi^{i})\,
e\,^{-\frac{i}{\hbar}\int d^{3}xdt\,\xi(\vec{x},t) P_{N}(\vec{x},t)}e\,^{-\frac{i}{\hbar}\int d^{3}xdt\,\xi^{i}(\vec{x},t) P_{N^{i}}(\vec{x},t)}\,,
\]
where we recasted the functional dependence on $\xi$ and on $\xi^{i}$, by introducing the functionals $\mathcal{K}(\xi)$ and $\mathcal{K}_{i}(\xi^{i})$, in place of $\mathcal{C}\left(\xi,\xi^{i}\right)$.
Then, we consider the following two different cases:
\begin{itemize}
\item As soon as we require the constance of the functionals $\mathcal{K}$ and $\mathcal{K_{i}}$,
the above integral become a delta-functional on the variable $P_{N}$ and $P_{N^{i}}$, apart from a normalization constant. 
Thus we get for $Z$
\begin{equation}
Z\left(J,W\right)=\mathcal{I}\,\int D(g_{a},P^{a})\,\delta\left(P_{N}\right)\,\delta(P_{N^{i}})
\,e^{\frac{i}{\hbar}S_{0}+i\int\left(g_{a}J^{a}+P^{a}W_{a}\right)}\,.
\label{deltapn}
\end{equation}
being 
\begin{equation*}
S_{0}=\int d^{3}xdt\left[p^{j}_{i}\;\partial_{t}u^{i}_{j}+p_{\alpha}\partial_{t}\alpha+ P_{N}\partial_{t}N+
P_{N^{i}}\partial_{t}N^{i}- N\mathcal{H} - N^{i}\mathcal{H}_{i}\right]\,.
\end{equation*}
Evaluating the delta-functionals, we obtain $Z$, calculated for the usual gravitational action $S_{HE}$, (\ref{s}) ,corresponding to vanishing momenta $P_{N}$ and $P_{N^{i}}$, and to a stationary evolution for the field, in the Schr\"odinger formulation, i.e. 
\begin{equation*}
Z\left(J,W\right)=\mathcal{I}\,\int D(g_{a},P^{a})\,
\,e^{\frac{i}{\hbar}S_{HE}+i\int\left(g_{a}J^{a}+P^{a}W_{a}\right)}\,.
\end{equation*}
Because of the constance of $\mathcal{K}(\xi)$ as a consequence of the delta term, we also get the following result for the T-product function; below we treat separately the quantities $\xi$ and $\xi^{i}$ since they enter in an equivalent way in the problem:
\begin{eqnarray*}
&&\left\langle 0\right|T\left(\widehat{P}_{N}(x_{1})...\widehat{P}_{N}(x_{n})\right)\left|0\right\rangle=
\frac{1}{i^{n}}\frac{\delta^{n}Z\left(J,W\right)}{\delta W_{N}(x_{n})...\delta W_{N}(x_{1})}\Bigg{|}_{J=0}\\[0.2cm]
&&=\mathcal{I}\,\int D(g_{a},P^{a})\,\delta\left(P_{N}\right)\,\delta\left(P_{N^{i}}\right)
P_{N}(x_{1})...P_{N}(x_{n})\,\,e^{\frac{i}{\hbar}S_{0}}=0\,.
\end{eqnarray*}
Obviously, the same arguments are valid for T-product functions containing $P_{N^{i}}$, if we take $\mathcal{K}_{i}(\xi^{i})$ as a constant.\\ 
We can conclude, that having a constant $\mathcal{C}$-term is equivalent to implement the primary constraints in their strong form and, therefore, to restate the WDE dynamics \cite{Har}, \cite{Har1}, \cite{HH}.

\item
Now, we will see that a different choice for $\mathcal{K}(\xi)$ and $\mathcal{K}_{i}(\xi^{i})$, in detail a Gaussian form, is equivalent to interpret the primary constraints in a relaxed way.\\
We start again from the expression (\ref{deltapn}) and we perform a constraint weakening by applying the standard procedure of the generating functional approach \cite{Tes}.\\
More precisely, in order to weak the constraint associated to the momentum $P_{N}$, we turn the $P_{N}$-delta, centered on the null function, into a $\lambda$-centered one. Furthermore, we introduce a Gaussian weight in the $\lambda$-parameter function, the corresponding normalization associated to this term being included into the new constant $\mathcal{I\,'}$. 
We show this method only for $P_{N}$, but we can extend it to all the other constraints associated to $P_{N^{i}}$.
This scheme yields the following generating functional:
\begin{eqnarray*}
Z\left(J,W\right)&=&\mathcal{I\,'}\,\int D(g_{a},P^{a})\,\delta\left(P_{N}-\lambda\right)\,\delta\left(P_{N^{i}}\right)
\,e^{\frac{i}{\hbar}S_{0}+i\int\left(g_{a}J^{a}+P^{a}W_{a}\right)}\\[0.2cm]
&&\qquad\left(\int \delta\lambda\,\,e^{-\frac{1}{2\sigma}\int d^{4}x\lambda^{2}(x)}\right)\,.
\end{eqnarray*}
Rewriting delta-functionals as integrals, we get
\begin{eqnarray*}
Z\left(J,W\right)&=&\mathcal{I\,'}\,\int D(g_{a},P^{a})\,\delta\xi\,\delta\xi^{i}\,\,
e^{-\frac{i}{\hbar}\int d^{4}x\,\xi\,(P_{N}-\lambda)}e^{-\frac{i}{\hbar}\int d^{4}x\,\xi^{i}\, P_{N^{i}}}\\[0.2cm]
&&\qquad e^{\frac{i}{\hbar}S_{0}+i\int\left(g_{a}J^{a}+P^{a}W_{a}\right)}
\left(\int \delta\lambda\,\,e^{-\frac{1}{2\sigma}\int d^{4}x\,\lambda^{2}(x)}\right)\,.
\end{eqnarray*}
Hence, we can identify the $\mathcal{K}(\xi)$ expression as:
\[\mathcal{K}(\xi)=\int \delta\lambda\,\, e\,^{\int d^{4}x\,\frac{i}{\hbar}\xi\,\lambda}\,
e^{-\frac{1}{2\sigma}\lambda^{2}}\,;
\]
by simple algebra, making the following substitution
\[\overline{\lambda}\equiv\left(\frac{1}{\sqrt{2\sigma}}\lambda+\xi\frac{i}{\hbar}\sqrt{\frac{\sigma}{2}}\right)
\]
and performing the $\overline{\lambda}$-integral, which result to be independent on $\xi$, finally we get
\[\mathcal{K}(\xi)=\mathcal{A}\, e\,^{\int d^{4}x\,\left(-\frac{\sigma}{2\hbar^{2}}\right)\,\xi^{2}}\,,
\]
being $\mathcal{A}$ a normalization constant.\\
From this formula, we can infer how, having a $P_{N}$-delta not centered in zero, so that $P_{N}$ is not vanishing, leads to a Gaussian form for the functional $\mathcal{K}(\xi)$.
In the limit in which the dispersion $\sqrt{\sigma}$ verge to zero, we obtain again the constance of $\mathcal{K}(\xi)$ and so the classic constraints. We can get the same result for $\mathcal{K}_{i}(\xi^{i})$.\\
As a consequence of this relaxation now we should get non-vanishing T-products of operators $\widehat{P}_{N}$ and $\widehat{P}_{N^{i}}$; we can show this, starting from a generating functional obtained by
assigning to $\mathcal{K}(\xi)$ and $\mathcal{K}_{i}(\xi^{i})$ a Gaussian form, as above, i.e.
\[\left\{\begin{array}{lll}
\mathcal{K}(\xi)&\propto&e^{-\frac{1}{2A}\int d^{4}x\,\xi^{2}(x)}\\[0.2cm]
\mathcal{K}_{i}(\xi^{i})&\propto&e^{-\left(\frac{1}{2A}^{3}\right)\Sigma_{i}\int d^{4}x 
\,(\xi^{i}(x))^{2}}\,,
\end{array}\right.
\]
where $A\equiv \frac{\hbar^{2}}{\sigma}$.\\
Dealing again with $\xi$-component only, we have
\[Z\left(J,W\right)=\int D(g_{a},P^{a})\delta\xi\,\delta\xi^{i}\,\,\mathcal{K}_{i}(\xi^{i})
\,\left(e^{-\frac{1}{2A}\int d^{3}x \,dt\xi^{2}(x)}\right)
e^{\frac{i}{\hbar}S+i\int\left(g_{a}J^{a}+P^{a}W_{a}\right)}\,,
\]
and, through this generating functional, we can write the T-products as
\begin{eqnarray*}
T_{1...n}&\equiv&\left\langle 0\right|T\left(\widehat{P}_{N}(x_{1})...\widehat{P}_{N}(x_{n})\right)\left|0\right\rangle\\[0.2cm]
&=&\mathcal{I\,'}\,\int D(g_{a},P^{a})\,\delta\xi\delta\xi^{i}\,\big[P_{N}(x_{1})...P_{N}(x_{n})\big]\,
e^{-\frac{1}{2A}\left(\int d^{3}x \,dt\,\xi^{2}(x)\right)}\,e^{\frac{i}{\hbar}S}\,.
\end{eqnarray*}
Because of $\xi$-term in the action, we can express every $P_{N}$ as a functional derivative in $\xi$, acting on the action exponential, so that the functions $T_{1...n}$ become
\[T_{1...n}=\frac{\mathcal{I\,'}}{A^{n}}\int D(g_{a},P^{a})\,\delta\xi\delta\xi^{i}\,\big[\xi(x_{1})...\xi(x_{n})\big]
e^{-\frac{1}{2A}\left(\int d^{3}x\,dt\,\xi^{2}(x)\right)}
e^{\frac{i}{\hbar}S}\,;\]
in this integral, we point out the $\xi$-term, i.e.
\[E(P_{N})\equiv \int \delta\xi\,e^{-\frac{i}{\hbar}\left(\int d^{3}x\,dt\,\xi(x)P_{N}(x)\right)}
e^{-\frac{1}{2A}\left(\int d^{4}x\,\xi^{2}(x)\right)}
\big[\xi(x_{1})...\xi(x_{n})\big]\,.
\]
If we turn it into
\[E(P_{N})=e^{\frac{A}{2\hbar^{2}}\,P^{2}_{N}} 
\int \delta\xi\, \left[\xi(x_{1})...\xi(x_{n})\right]e^{-\frac{1}{2A}\int d^{3}x\,dt
\left(\xi+\frac{iA}{\hbar}P_{N}\right)^{2}}\,,
\]
we can see that it is not vanishing, being a Gaussian integral not null-centered, though around an imaginary value.\\
As a consequence, the whole expression of the T-products $T_{1...n}$ no longer vanish and this is the expected quantum relaxation of the primary constraints.
\end{itemize}

\section{Evolutionary quantum gravity.}
The next step, in our work, is to show the dynamical consequences, in the Schr\"odinger approach, 
of the generating functional resulting from the Gaussian choice for $\mathcal{C}(\xi,\xi^{i})$.\\
In this representation, having non-vanishing T-products $T_{1...n}$ of operators $\widehat{P}_{N}$ and $\widehat{P}_{N^{i}}$, corresponds to require the following condition on $\psi$:
\[\widehat{P}_{N}\psi\neq0\,,\qquad\widehat{P}_{N^{i}}\psi\neq0\,.
\]
As in the canonical formulation, we have $\widehat{P}_{N}=-i\hbar\frac{\delta}{\delta N}$ and $\widehat{P}_{N^{i}}=-i\hbar\frac{\delta}{\delta N^{i}}$, so that now the wave functional $\psi$ has to depend
on the the slicing structure, that is on $N$ and $N^{i}$. Moreover, via the Hamilton equations
(extended to the operator form), the violation of the secondary constraints follows directly, i.e.
\[\widehat{\mathcal{H}}\psi\neq0\,, \qquad \widehat{\mathcal{H}}_{i}\psi\neq0\,,
\]
as well as the explicit breakdown of the quantum theory invariance, under 4-diffeomorphisms.\\
To continue our analysis, we point out that:
\begin{itemize}
\item The dependence of $\psi$ on the variables $N$ and $N^{i}$ can be seen as a 
dependence on the label time $t$.
Reminding the definition of the deformation vector (\ref{defo}),
we can refer to the lapse function and shift vector dependence, as to that one on $y\,^{\mu}$, the field  
which fixes the splitting; thus, we can write the temporal derivative as
\[\partial_{t}\psi\equiv\int d^{3}x\, \frac{\delta\psi}{\delta y^{\mu}}\,\partial_{t}y^{\mu}.
\]
As we have seen, because of the relaxation, we get
\[\frac{\delta\psi}{\delta y^{\mu}}\neq0\Longrightarrow 
\partial_{t}\psi\neq0
\]
and so the $\psi$ temporal dependence outcomes.
\item In the Schr\"odinger representation, we can express the wave functional $\psi$ as a path integral, 
which is also a transition amplitude between two states, i.e. the propagator \cite{FH}, \cite{Sak}.\\
(For an extension of path integral approach to the General Relativity see also \cite{Haw}).\\
The metric configurations $g_{a},g_{0}$ (corresponding to different times, $t$ and $t_{0}$, and to associated hypersurfaces $\Sigma^{3}$ and $\Sigma^{3}_{0}$) characterize the final and the initial state for the transition.\\ 
In agreement to the analysis of the previous section, we assign the following path integral, which correspnds 
to the adopted scheme of constraints relaxation: 
\begin{equation}
\left\langle g_{b},t\,\Big{|}\,g_{a},t_{0}\right\rangle=\int D(g_{a},P^{a})\,\delta\xi\,\delta\xi^{i}\,\mathcal{C}(\xi,\xi^{i})\,e^{\frac{i}{\hbar}S}\,,
\label{pro}
\end{equation}
where $\mathcal{C}(\xi,\xi^{i})$ has to be thought Gaussian (indeed, the below analysis holds even
for a non-Gaussian form of $\mathcal{C}$, but it is constant)\\
This integral does not contain boundary terms, because the hypersurfaces are compact and boundaryless.\\ 
We remind that $g_{a}$ stands for the metric configuration, in the time t, given by the variables 
$(u,\alpha,N,N^{i})$, as well as $P^{a}$ stands for $(p,P_{\alpha},P_{N},P_{N^{i}})$.
\end{itemize}

Starting from (\ref{pro}), we rewrite the corresponding gravitational action as
\[S=\int d^{3}x \,P^{a}D(g_{a}) -\int \left(H_{0} +H_{(\xi,\xi^{i})}\right)dt
\]
Here, we have pointed out the term $H_{\xi,\xi^{i}}$, having the form
\[H_{(\xi,\xi\,^{i})}\equiv\int d^{3}x\left(\xi P_{N}+\xi^{i}P_{N^{i}}\right)\,;
\]
which is reabsorbed by the integration on $\xi$ e $\xi^{i}$, as follows:
\[\left\langle g_{a},t\,\Big{|}\,g_{0},t_{0}\right\rangle=
\int D(g_{a},P^{a})\,\mathcal{F}(P_{N},P_{N^{i}})\,e^{\frac{i}{\hbar}S_{0}}\,;
\]
$\mathcal{F}(P_{N},P_{N^{i}})$ is a functional depending on the momenta $P_{N}$ and $P_{N^{i}}$, resulting from the above integration.\\
Now, we can obtain the dynamical equation, taking
\begin{eqnarray*}
i\hbar\partial_{t}\psi=i\hbar\partial_{t}\left\langle \,g_{a},t\,\Big{|}\,g_{0},t_{0}\right\rangle
&=&\int D(g_{a},P^{a})\,\mathcal{F}(P_{N},P_{N^{i}})\,(-\partial_{t}S_{0})\,\,e\,^{\frac{i}{\hbar}S_{0}}\\[0.2cm]
&=&\int D(g_{a},P^{a})\,\mathcal{F}(P_{N},P_{N^{i}})\,H\left(g_{a},P^{a}\right)\,\,e\,^{\frac{i}{\hbar}S_{0}}\\[0.2cm]
&=&\int D(g_{a},P^{a})\,\mathcal{F}(P_{N},P_{N^{i}})\,:\widehat{H}:\,e\,^{\frac{i}{\hbar}S_{0}}\,,
\end{eqnarray*}
where explicitly $\psi\equiv\psi(g_{a},t)$.\\
Above, in the last passage, we have changed the super-Hamiltonian and the super-momentum into operators,
disregarding higher power in $\hbar$, as follows:
\begin{eqnarray*}
\mathcal{H}\left(g_{a},\frac{\delta S_{0}}{\delta g_{a}}\right)\,e^{\frac{i}{\hbar}S_{0}}&=&:\widehat{\mathcal{H}}
\left(u,\alpha,-i\hbar\frac{\delta}{\delta u},-i\hbar\frac{\delta}{\delta \alpha}\right):
e^{\frac{i}{\hbar}:S_{0}:}\,,\\[0.2cm]
\mathcal{H}_{i}\left(g_{a},\frac{\delta S_{0}}{\delta g_{a}}\right)e^{\frac{i}{\hbar}S_{0}}&=&:\widehat{\mathcal{H}}_{i}
\left(u,\alpha,-i\hbar\frac{\delta}{\delta u},-i\hbar\frac{\delta}{\delta \alpha}\right):e^{\frac{i}{\hbar}:S_{0}:}\,,
\end{eqnarray*}
where we used the Hamilton-Jacobi relation between variables and conjugate momenta 
$P^{a}=\frac{\delta S_{0}}{\delta g_{a}}$. \\
At the end, if we remark that the functional derivatives contained in $:\widehat{H}:$ commute with $\mathcal{F}$ and with the integration measure, we get, in the limit of small $\hbar$, the Schr\"odinger equation for the gravitational quantum field
\begin{equation}
i\hbar\,\partial_{t}\psi=\,:\widehat{H}:\psi\label{scho}\,.
\end{equation}
In the derivation of this evolutionary approach to quantum gravity, we need not a specific form for $\mathcal{C}$, but requiring it is not a constant functional. Indeed, it is possible to show that having a non-constant $\mathcal{C}$
corresponds to a non-vanishing vacuum expectation value of the momenta $\widehat{P}_N$ and $\widehat{P}_{N^{i}}$, i.e. $\left\langle 0\right|P_N\left|0\right\rangle\neq0$ and  $\left\langle 0\right|P_{N^{i}}\left|0\right\rangle\neq0$.\\
This issue indicates that the existence of an evolutionary quantum gravity does not require that the primary constraints are violated in a strong form 
(i.e. all the T-products for $\widehat{P}_N$ and $\widehat{P}_{N^{i}}$ vanish).\\
By other words, to get a Schr\"odinger dynamics, it is sufficient to break the primary constraints (and therefore the secondary ones too) only in terms of their expectation values (see also \cite{Nico}).

\section{Classical limit of the theory and the Eckart fluid.}
In this section, we want to show how the dynamical equation (\ref{scho}), founded for the gravitational field through the relaxation, implies the presence of a particular fluid in the system, when taking the classical limit. Such a fluid results to have the Eckart fluid characteristics \cite{Eck}. (For other approaches, which correlates the presence of a fluid in the dynamics to the appearance of a time in quantum gravity, see \cite{Kuc4}, \cite{Kuc5}, \cite{KuT}, \cite{Bro} \cite{1a}, \cite{1b},  \cite{1c}, \cite{1d}, \cite{1e}, \cite{1f}, \cite{Mon}).\\
Let us consider the following development for the wave functional $\psi$, over the 
super-Hamiltonian and super-momentum eigenfunctions
\begin{equation}
\psi(u^{n}_{j},\alpha,N,N^{i},t)=\int D\omega Dk_{i} \,\chi(\omega,k_{i})\,e^{-\frac{i}{\hbar}(t-t_{0})\,
\int d^{3}x\left[N\omega+N^{i}k_{i}\right]}\,;
\label{kappa}
\end{equation} 
$\omega\equiv\omega(\vec{x})$ and $k_{i}\equiv k_{i}(\vec{x})$ are the eigenvalues functions, and 
\[\chi(\omega,k_{i})=\chi(u^{a}_{b},\alpha,\omega,k_{i})
\]
is the corresponding eigenfunctional, i.e. we get
\begin{eqnarray}
\widehat{\mathcal{H}}\,\chi(\omega,k_{i})&=&\omega\,\chi(\omega,k_{i})\,,\label{om}\\[0.2cm]
\widehat{\mathcal{H}}_{i}\,\chi(\omega,k_{i})&=&k_{i}\,\chi(\omega,k_{i})\,.\label{ka}
\end{eqnarray}
Now, we mean to take the classical limit of these eigenvalues problems, via a WKB approach, which provide us the following form for $\chi(\omega,k_{i})$, to be inserted into (\ref{om}) and (\ref{ka}):
\[\chi(\omega,k_{i})=\mathcal{M}\,\,e\,^{\frac{i}{\hbar}\mathcal{P}}
\]
being $\mathcal{M}$ and $\mathcal{P}$ the modulus and the phase of $\chi$, respectively.\\
In the limit $\hbar\longrightarrow0$, the substitution result 
yields the gravitational Hamilton-Jacobi equation and its super-momentum equivalent, 
containing two additional terms, which correspond to the classical limit of $\omega$ and $k_{i}$\,
i.e. $\overline{\omega}$ and $\overline{k}_{i}$ respectively:
\begin{eqnarray}
\mathcal{H}\left(g_{a},\frac{\delta\mathcal{P}}{\delta g_{a}}\right)=\overline{\omega}\,,\label{wkbom}\\[0.2cm]
\mathcal{H}_{i}\left(g_{a},\frac{\delta\mathcal{P}}{\delta g_{a}}\right)=\overline{k}_{i}\,.\label{wkbk}
\end{eqnarray}
We recast this Hamilton-Jacobi system of equations into the Einstein one.\\ 
Using well-known results \cite{Thi}, \cite{Wal}, it is possible to show 
that the following relations hold:
\begin{enumerate}
\item $G_{\mu\nu}n^{\mu}n^{\nu}=XT_{\mu\nu}n^{\mu}n^{\nu}=-\frac{\mathcal{H}}{2\sqrt{g}}
=-\frac{\overline{\omega}}{2\sqrt{g}}$
\item $G_{\mu\nu}\partial_{i}y^{\mu}n^{\nu}=XT_{\mu\nu}\partial_{i}y^{\mu}n^{\nu}=\frac{\mathcal{H}_{i}}{2\sqrt{g}}
=\frac{\overline{k}_{i}}{2\sqrt{g}}$
\item $G_{\mu\nu}\partial_{i}y^{\mu}\,\partial_{j}y^{\nu}=XT_{\mu\nu}\partial_{i}y^{\mu}\,\partial_{j}y^{\nu}=0$
\end{enumerate}
where $X$ denotes the Einstein constant.\\
Taking into account these three relations, we arrive to construct the energy-momentum tensor, 
associated to the classical limit of the eigenvalues problems (\ref{wkbom}) and (\ref{wkbk}),
\[T_{\mu\nu}=\epsilon\,n_{\mu}n_{\nu}+2s_{(\mu}n_{\nu)}\,;
\]
the identifications
\begin{eqnarray*}
\epsilon&=&\frac{-\overline{\omega}}{2X\sqrt{g}}\,,\\[0.2cm]     
s_{\mu}&=&-\frac{\overline{k}\,^{i}}{2X\sqrt{g}}\,\partial_{i}y^{\rho}\,h_{\rho\mu}\,,
\end{eqnarray*}
take place.\\
In particular the above third relation assure us the tensor $T_{\mu\nu}$ does not contain any pure spatial component.\\

The so founded energy-momentum tensor $T_{\mu\nu}$ has the expression of that one associated to an Eckart fluid, as soon as we identify $\epsilon$ as the fluid energy density, $n^{\mu}$ as its 4-velocity and $s_{\mu}$ as its thermal conduction vector \cite{Eck}.\\
This fluid is co-moving with the spatial hypersurface and contains a time-like vector corresponding to its 4-velocity, so that its presence gives a physical meaning to the label time $t$, used in the splitting procedure.
We can infer that, in quantum gravity, the existence of a time and the presence of such a fluid in the gravitational dynamics are both aspects of the same physical entity.

\section{Concluding remarks.}
The main issue of our analysis has to be regarded the construction of a phase-space generating functional, for an evolutionary quantum gravity. We have shown how the quantum implementation of the primary constraints can be controlled through the form taken by a free functional of the associated Lagrangian multipliers. When this functional is a constant one, the implementation of the constraints takes place in its strong form (the operators $\widehat{P}_{N}$ and $\widehat{P}_{N^{i}}$ vanish identically), so that the standard Wheeler-DeWitt approach is reproduced.
But, if we address Gaussian functional form on $\xi$ and $\xi^{i}$, then, the resulting dynamics corresponds to the relaxation of the primary constraints, i.e. the T-products associated to $\widehat{P}_{N}$ and $\widehat{P}_{N^{i}}$ are different from zero, for any number of points. \\
The physical meaning of the time variable, we have introduced in taking the T-products (namely the label time), is recognized as soon as we extend the relaxation scheme of the primary constraints to the path-integral formulation of the Shr\"odinger picture.\\
In fact, in this picture, it is enough to require that the free functional, on $\xi$ and $\xi^{i}$, has a non-constant form, to get an evolutionary quantum dynamics, in the limit of small $\hbar$, as described by the Schr\"odinger equation.
Having a non-constant functional ensures only that $\left\langle0\right|\widehat{P}_{N}\left|0\right\rangle\neq0$ and
$\left\langle0\right|\widehat{P}_{N^{i}}\left|0\right\rangle\neq0$, but it does not imply the vanishing behavior of any T-product function. We can infer that, in the limit of small $\hbar$, the appearance of a time in quantum gravity is associated to weaker relaxation of the primary constraints. Indeed, the physical meaning of the label time outcomes when taking the classical limit of the Schr\"odinger equation.\\ 
Summarizing, we started from the classical Hamiltonian dynamics and assumed the label time suitable for constructing T-products functions. The classical limit for the theory provided the appearance of an Eckart fluid within the Einstein equations, as required by the selfconsistence of the spacetime splitting and by the existence of a real time variable. Thus, we have shown that if we have a physical time in quantum level, it results into a reference fluid in the classical behavior.

\end{document}